\documentclass[preprint,prb,showpacs,preprintnumbers,amsmath,amssymb]{revtex4}
\usepackage{epsfig,amsmath,graphicx,amssymb}
\usepackage{graphicx}
\usepackage{amsmath}
\usepackage{float}
\usepackage{color}
\begin{document}
\title{
Synchronous Characterization of Semiconductor Microcavity Laser Beam
}

\author{Tao Wang$^{\rm 1,2,a}$ and Gian-Luca Lippi$^{\rm 1,2}$}
\affiliation{
$^{\rm 1}$ \mbox{Institut Non Lin\'eaire de Nice, Universit\'e de Nice Sophia Antipolis}\\
$^{\rm 2}$ \mbox{CNRS, UMR 7335, 1361 Route des Lucioles, F-06560 Valbonne, France}\\
$^{\rm a}$ email:  taowang166@gmail.com}

\date{\today}

\begin{abstract}
We report on a high-resolution double-channel imaging method used to synchronously map the intensity- and optical-frequency-distribution of a laser beam in the plane orthogonal to the propagation direction. The synchronous measurement allows us to show that the laser frequency is an inhomogeneous distribution below threshold, but that it becomes homogeneous across the fundamental Gaussian mode above threshold. The beam's tails deviations from the Gaussian shape, however, are accompanied by sizeable fluctuations in the laser wavelength, possibly deriving from manufacturing details and from the influence of spontaneous emission in the very low intensity wings. In addition to the synchronous spatial characterization, a temporal analysis at any given point in the beam cross-section is carried out. Using this method, the beam homogeneity and spatial shape, energy density, energy center and the defects-related spectrum can also be extracted from these high-resolution pictures.
\end{abstract}

\maketitle

\section{Introduction}
Lasers have been one of the most versatile sources of electromagnetic energy both for scientific research and for innumerable practical applications~\cite{Demtroder2009,Duarte2009,Sparkes2008}.  Thus, the control of the laser beam parameters has become of paramount importance~\cite{Gavrilovic1995,Sheikh2011} and a large effort has been devoted to beam characterization. In the past several decades, laser beams have been widely used in different fields both for diagnostics and as a powerful, precise and clean (non-contact) tool to modify the structure of matter.  Their applications range from materials processing~\cite{Duocastella2012}, lithography~\cite{Gan2013}, medical treatments~\cite{Schulze2010}, laser printing~\cite{Kataoka2001}, optical data storage~\cite{Gan2013}, micro-machining ~\cite{Holmes2006} in the electronics industry, isotope separation~\cite{Parvin2004}, optical processing~\cite{Minoshima2001,Ready2001,Steen2010}, optical manipulation~\cite{Ashkin2006} and laboratory research~\cite{Dickey2005}.  All these fields of application require a high degree of control on the laser source for optimal guiding and shaping, thus control on the process.  In addition, a good characterization of the beam properties~\cite{Siegman1998} can provide information on the quality of fabrication, while continuously monitoring the beam shape during processing may be necessary for applications where beam-degradation-induced defects are an issue~\cite{Roundy}. Although characterizing the laser beam has been a topic of study and discussion for over two decades~\cite{Weber1992,Siegman1993}, there still are aspects of the problem which can benefit from new advancements.

Typically, the techniques developed to characterize a laser beam provide information about the energy distribution~\cite{Alsultanny2006,Gebert2014} or the integrated frequency spectrum ~\cite{Look2010,Liu2011} independently of each other. The simultaneous measurement of the two quantities, which allows for an analysis of the spatially resolved spectral features (frequency purity, linewidth, homogeneous vs. inhomogeneous broadening, as well as spatio-temporal coherence), is not readily available.  We propose here a characterization method which allows for the simultaneous measurement of the energy distribution and the emission frequency in the transverse plane.  The single mode microcavity laser beam, which we use as an example in Section~\ref{expt}, is characterized using the intensity and frequency distribution. However, extensions to additional channels (e.g., monitoring the polarization of the e.m. field or the local temporal dynamics) are readily obtained from the setup by using multiple-way splitters or by cascading simple splitters.  This technique allows us to quantify the beam quality, as well as to monitor the actual beam profile, while at the same time following the spectral composition of the radiation emitted at each point.  We will illustrate the technique by analysing the mode emitted by a semiconductor laser, showing that some nontrivial information can be recovered in this way (e.g., beam homogeneity, energy center, emission frequency and frequency stability).

\section{Instrument}

The scope of this instrument is to provide a high-resolution sampling of the transverse field distribution issued by a source for characterization through multiple, simultaneous analysis channels.  Three different components constitute the instrument:  the mechanics, which has to ensure good positioning in the transverse plane, but also a good  control on the orientation of the fiber tip which samples the radiated field; the optics -- a fiber suitably split to provide the multiple outputs (either through N-way splitters or cascaded 2-way splitters); the control -- a computer controlled interface driving the positioning of the fiber tip.  Finally, additional instrumentation is used to shape the beam to study (e.g. for choosing the plane to sample) and for the analysis.  This ensemble is interchangeable and depends on the kind of measurement being performed.

\subsection{Mechanics}

\begin{figure}[H]
\centering
\includegraphics[width=0.45\linewidth,clip=true]{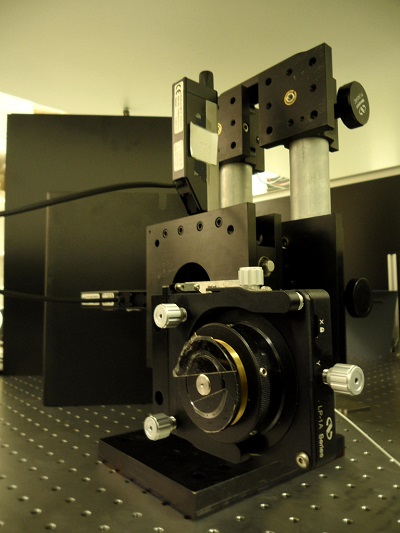}
\includegraphics[width=0.45\linewidth,clip=true]{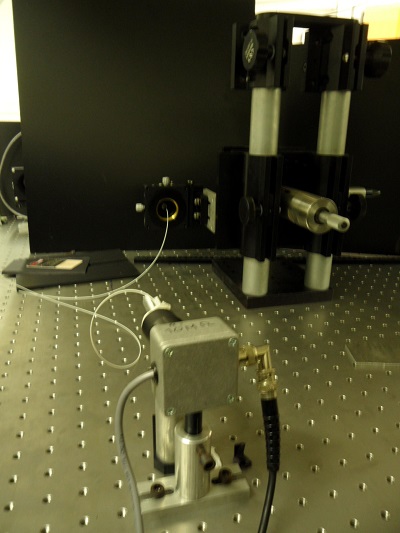}
\caption{Left panel:  front view of the mechanical parts of the instrument -- the optical fiber is held in the center of the Newport LP-1A five axis translation stage; the two motorized screws are also visible.  Right panel:  back view of the mechanical setup -- the fiber (white) exiting the mechanical holder and arriving to the detector (bottom left part of the figure) is recognizable, as well as the long-travel micrometric z-translation.}
\label{photos}
\end{figure}

The mechanical parts are assembled from commercial, precision parts which allow for accurate positioning of a fiber tip, scanned across the optical field to be sampled.  A five-axis positioner (LP-1A, Newport) is mounted on a computer-controlled two-axis translation stage (Newport 405) fitted with precision, long travel, high-speed motorized actuators (Newport LTA-HS) with minimum incremental motion $0.1 \mu m$ and (unidirectional) repeatability $0.5 \mu m$.  The total travel range of the mounted system is approximately $1 cm$ in each direction.  The two translation stages are mounted vertically, and orthogonally to each other to provide $x-y$ scans, on a sturdy support (standard plates mounted on damped Newport rods).  Home-built adapters are used to mount the fiber in the center of the five-axis positioner, to support the latter onto the translation stages and to hold the ensemble on the base. An additional, long-travel micrometric z-displacement is added onto the vertical holder to provide a long range of adjustment along the propagation direction of the optical radiation.  Fig.~\ref{photos} show two images of the mechanical parts.

\subsection{Optics}

The optical sampling is obtained by scanning a fiber across the transverse intensity distribution.  Single or multi-mode fibers can be used for this purpose, depending on the application.  In this realization, we use a single-mode fiber.  

Specifically, one arm of a single mode (at $\lambda = 980 nm$)  fiber splitter ($5.8 \mu m$ core diameter, Thorlabs FC980-50B-FC, 50:50) is mounted on axis in the LP-1A positioner, so as to profit of all the possible adjustments:  the fiber can be manually positioned ($0.75 \mu m$ sensitivity) on the maximum of the beam profile in the transverse plane (e.g., before starting a scan) and finely adjusted in the longitudinal direction ($z$-adjustment with $1 \mu m$ sensitivity), e.g., to place it in the focal plane of an imaging system (for near-field measurements).  In addition, it is possible to tilt the fiber tip to optimize coupling from the incoming radiation (and to avoid feedback into the laser) using the two angular adjustments ($\theta$ and $\phi$:  sensitivity $2 \rm arcsec$). In the example we are providing, two optical channels are available and will be used to monitor the intensity profile and the optical spectrum in each point of the intensity distribution.

\subsubsection{Fiber selection}
Fiber selection plays an important role in this instrument and determines some of its key performances.  For instance, use of a polarization-maintaning fiber will allow for polarization analysis by preserving its state during guided propagation.  Similarly, choosing a single mode, rather than a multimode fiber determines the quality of the information.  The single mode fiber allows for better spatial resolution, during sampling, and higher fidelity in spectral analysis (depending also on the specifications of the chosen spectrum analyzer -- cf. below).  Multimode fibers (with diameters from tens to hundreds of micrometers) offer the advantage of carrying a much larger amount of light, even in spite of reduced spatial resolution (potential overlap between sampled points), but their incoupling conditions are sensitive to motion and the modal distribution of the guided light may be less reproducible when repositioning.  Furthermore, high-end spectrum analyzers are often specified with single-mode fibers (or with specific multimode ones).  Thus, when choosing the fiber one must take into account the characteristics of the instruments which are connected to it.  The main shortcoming of the single mode fiber is that very weak signals may fall below the detection threshold, due to its small diameter.   Otherwise, its features are the best suited for the analyzer we present here.

\subsubsection{Fiber coupling}
In order to maintain a correct coupling across the full beam, the angle of incidence of each wavevector (independently on its distance from the optical axis) must remain smaller than the angle of acceptance of the fiber.  A complete discussion is offered in Appendix A (Section~\ref{appendixA}).  Here we simply consider two limiting cases:  the near and far field.  

The near field is characterized by a planar wavefront since it is normally accessed by imaging the true near field of the source (as is the case of the VCSEL we use in the example of Section~\ref{expt}) and analyzing the beam in the image focal plane.  Thus, all wavevectors are parallel to the optical axis and satisfy the fiber coupling constraints once the instrument is aligned.  Errors in positioning result in small angles in the peripheral portions of the beam, as discussed in Appendix A (Section~\ref{appendixA}), and can be accordingly quantified.  

The far field can either be obtained by letting the beam propagate far (three or more Rayleigh lengths), or by using a lens to conjugate to infinity.  The latter solution is normally preferable as it allows for more compact setups and for a higher local intensity (for small lasers the beam divergence lowers the local intensity to the point where no measurements become possible).  The lens (or aspheric collimator, normally used for VCSELs) produces a (quasi) plane wave, thus automatically fulfilling the coupling condition, as in the case just presented.  As in the discussion of the near field, some deviation is introduced by the imperfections of the optical system which leave a residual divergence, to be compared to the angle of acceptance of the fiber.  In section~\ref{expt} we will specifically quantify both cases for the laser we use for demonstration.

\subsection{Positioning and Control}

\begin{figure}[H]
\centering
\includegraphics[width=0.45\linewidth,clip=true]{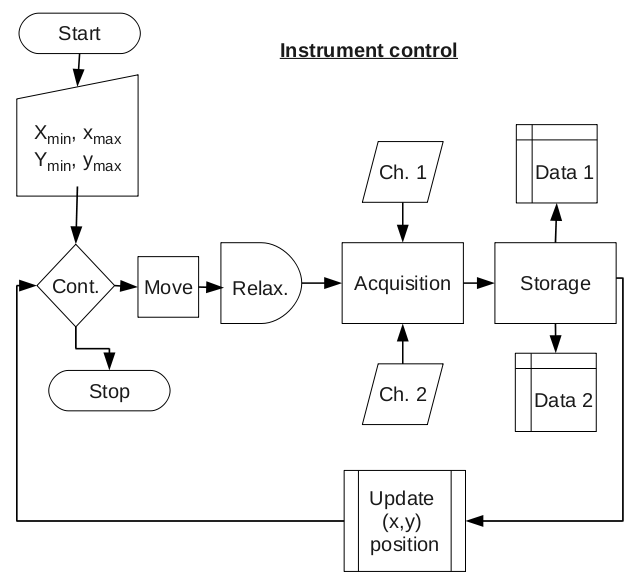}
\includegraphics[width=0.45\linewidth,clip=true]{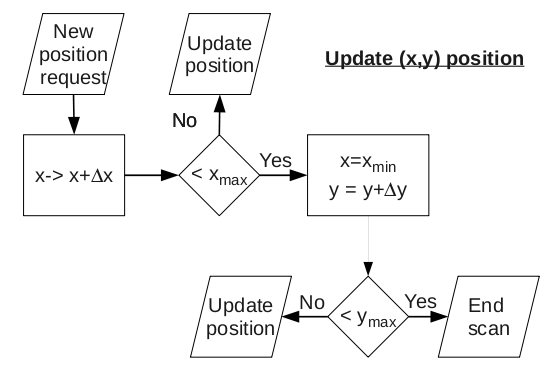}
\caption{Block diagram for the control procedure.  The right part of the figure (``Update (x,y) position") details the control block appearing with the same name on the left part.}
\label{principle}
\end{figure}

The positioning and control functionalities are explained in the flow diagram of Fig.~\ref{principle}.  The computer pilots the x and y translation stages to position the optical fiber at the first position of the acquisition matrix (top left).  After the position is attained, information arrives from the stage controls to the program signalling the success of the operation.  A pause ensues to allow for the relaxation of the mechanical vibrations of the fiber tip, then the measurement is enabled.  The signal is acquired from the measurement channels (e.g., two, as in the schematics) and is stored in the corresponding measurement matrices. The program updates the position of the fiber tip incrementing first along the row, then -- once the end of the row is attained -- passing to the next column, first position. The information is passed to the stage controls and the loop resumes until the full position matrix is complete.

The absolute values of the extreme positions in both directions is fixed by the operator before starting the procedure (compatibly with the values permitted by the material), together with the step both in x and y (which do not need to be equal). The number of points in the position matrix can be chosen by the operator prior to starting the measurement and is limited on the lower range by the resolution of the positioning system and the fiber tip size (e.g., to obtain images which do not overlap), on the upper range by the maximum excursion permitted by the translation stage and micro-positioning system.  In most cases, the maximum size is chosen to adapt to the field of view that is desired, typically smaller than the maximum range.  Reasonable scans cover matrices of $50 \times 50$ points, which give a good compromise between spatial resolution and acquisition time.

\section{Experiment}\label{expt}

\begin{figure}[H]
\centering
\includegraphics[width=0.65\linewidth,clip=true]{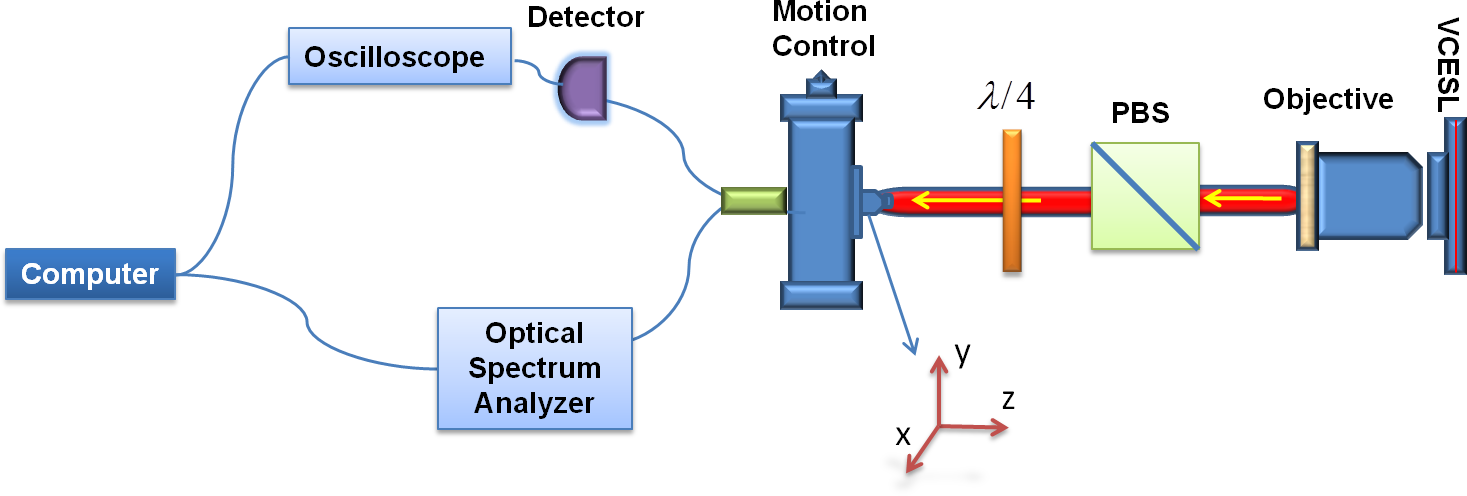}
\caption{Measurement system setup.  The microscope objective (x40) relays the near-field of the VCSEL emission to the entrance face of the single-mode fiber, mounted on the motion-control system.  The Polarizing Beam Splitter (PBS) and Quarter-Wave Plate ($\lambda/4$) form an optical isolator to minimize feedback into the laser.  The signal from the photodetector is digitized by an Agilent DSO-X 3024A oscilloscope.  Besides storing the data, the computer controls the whole acquisition process.}
\label{Experimentalsetup}
\end{figure}

The complete measurement system is illustrated in Fig.~\ref{Experimentalsetup}. 

The emitter we investigate is a single mode semiconductor Vertical-Cavity Surface-Emitting Laser (VCSEL) manufactured by ULM-Photonics and emitting at $\lambda = (980 \pm 3) nm$. Its maximum output power is 2 mW (at maximum current $i = 2 mA$) with thermal tunability $\frac{d\lambda}{d T} = 0.06 nm K^{-1}$.  The laser is mounted on a Thorlabs unit (TCLDM9) and is temperature-stabilized to better than $0.1 K$.  

In order to measure the near-field of the laser emission, we form its image with a microscope objective (x40, $NA = 0.65$) onto the analysis plane, where we position the fiber tip.  Interposed between microscope objective and fiber we place an optical isolator, to minimize feedback into the VCSEL, formed by a Polarizing Beam Splitter (PBS, Thorlabs PBS103) and a Quarter-Wave-Plate ($\lambda$/4, Thorlabs WPQ05M­980); the optical path length is compensated when positioning the fiber tip in the image plane of the microscope objective. Notice that the objective's numerical aperture is purposefully not matched to the laser's beam divergence~\cite{ULMPhot} because in the following we want to collect the spontaneous emission (thus also in angular directions escaping the laser beam diameter).

The laser light collected by the fiber is equally split between the photodetector (UDT-455, bandwidth $\approx 70 kHz$), and the optical spectrum analyzer (Agilent 86142B).  The digitized data (with 16 bit resolution) are stored in a computer which controls, through a Python interface, the whole acquisition system.  The scan of the transverse field intensity distribution is performed in steps of $6 \mu m$ over a total scanning range $(0.3 \times 0.3) mm^2$, thus producing two $50 \times 50$ output matrices containing the intensity and spectral information.  The time needed for positioning the fiber, letting the mechanical oscillations relax, acquiring the data and starting the new positioning takes in average $2 s$, thus requiring approximately $1^h \ 23^{\prime}$ for an entire image.  The resulting beam parameter stability issues are discussed in the following.  Notice that this setup is very flexible as it allows for the measurement of the laser's near field emission with the inclusion of a suitably adjusted microscope objective (cf. Fig.~\ref{Experimentalsetup}), but also of the far field (or any other intermediate plane) choosing an appropriate setup and selecting the corresponding scanning range and resolution.

As discussed in detail in Appendix A (Section~\ref{appendixA}) it is important to estimate the quality of beam coupling into the fiber across the whole measured pattern.  Using the far field divergence the VCSEL's beam waist can be estimated at $w_0 \approx 1.4 \mu m$ ($w_0 = \frac{\lambda}{\pi \theta_{ff}}$, as in~\cite{Siegman1986}, where $2 \theta_{ff} = 25^{\circ}$, according to the manufacturer's specifications~\cite{ULMPhot}), thus $z_R \approx 1 \times 10^{-2} m$ and $\theta_{ff} \approx 0.006 rad$ (having assumed a magnification x40 for the beam waist, given by the microscope objective).  With the angle of acceptance for the fiber specified by eq.~(\ref{defNA}), we can estimate the maximum tolerable error -- compatible with correct coupling to the fiber -- in the identification of the near-field plane (eq.~(\ref{Deltaz})) to be $\Delta z \approx 2 mm$, which is very large compared to the accuracy which can be attained with micropositioning.  Thus, we can rest assured that the coupling into the fiber correctly reproduces the field distribution, as confirmed by Fig.~\ref{beam, beamtopquan}, which properly reproduces even the far wings of the intensity distribution.

For the far-field, we ressort to the apparent residual divergence of the beam.  Measuring the propagation at two positions distant approximately 1m from each other, we find an increase in the beam radius of approximately 1mm.  Thus, the corresponding residual divergence is $\theta_d \approx 10^{-3} rad \ll \theta_a \approx 0.12$ (cf. eq.~(\ref{defNA})), which ensures again that the coupling into the fiber is correct across the full beam's cross section.

\section{Results and Discussion}
\begin{figure}[H]
\centering
\includegraphics[width=0.65\linewidth,clip=true]{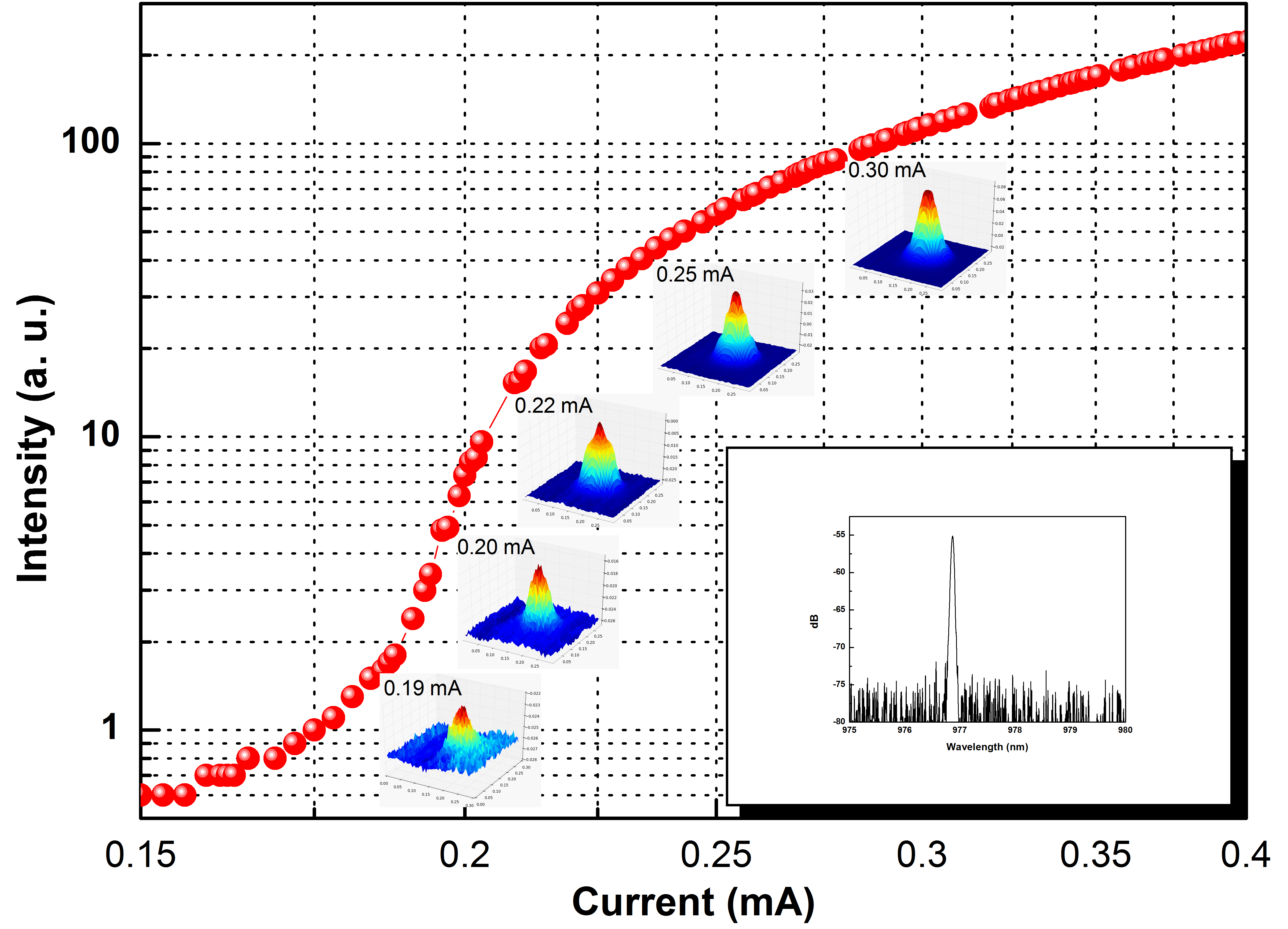}
\caption{Output intensity as a function of injection current. The color insets are the intensity distribution profiles at different pumping current. The inset shows the optical spectrum at $i = 0.30 mA$.}
\label{IVcurve}
\end{figure}

Fig.~\ref{IVcurve} shows the typical input-output laser characteristic curve with a threshold current $i_{th} \approx 0.2 mA$.  The optical spectrum (cf. inset, bottom right of Fig.~\ref{IVcurve})confirms single-mode operation at $i = 0.30 mA$, while a series of intensity distribution profiles, reconstructed from the photodetector information, describe the formation of Gaussian shape with injection current. 

The nearly-gaussian intensity profile at $i = 0.30 mA$ is shown in Fig.~\ref{beam, beamtopquan} (left), with the top view given in the right panel.  The red lines represent gaussian fits of axial intensity cuts in the x and y directions, respectively. The pictures indicate an almost symmetric TEM$_{00}$ mode profile.

\begin{figure}[H]
\centering
\includegraphics[width=0.45\linewidth,clip=true]{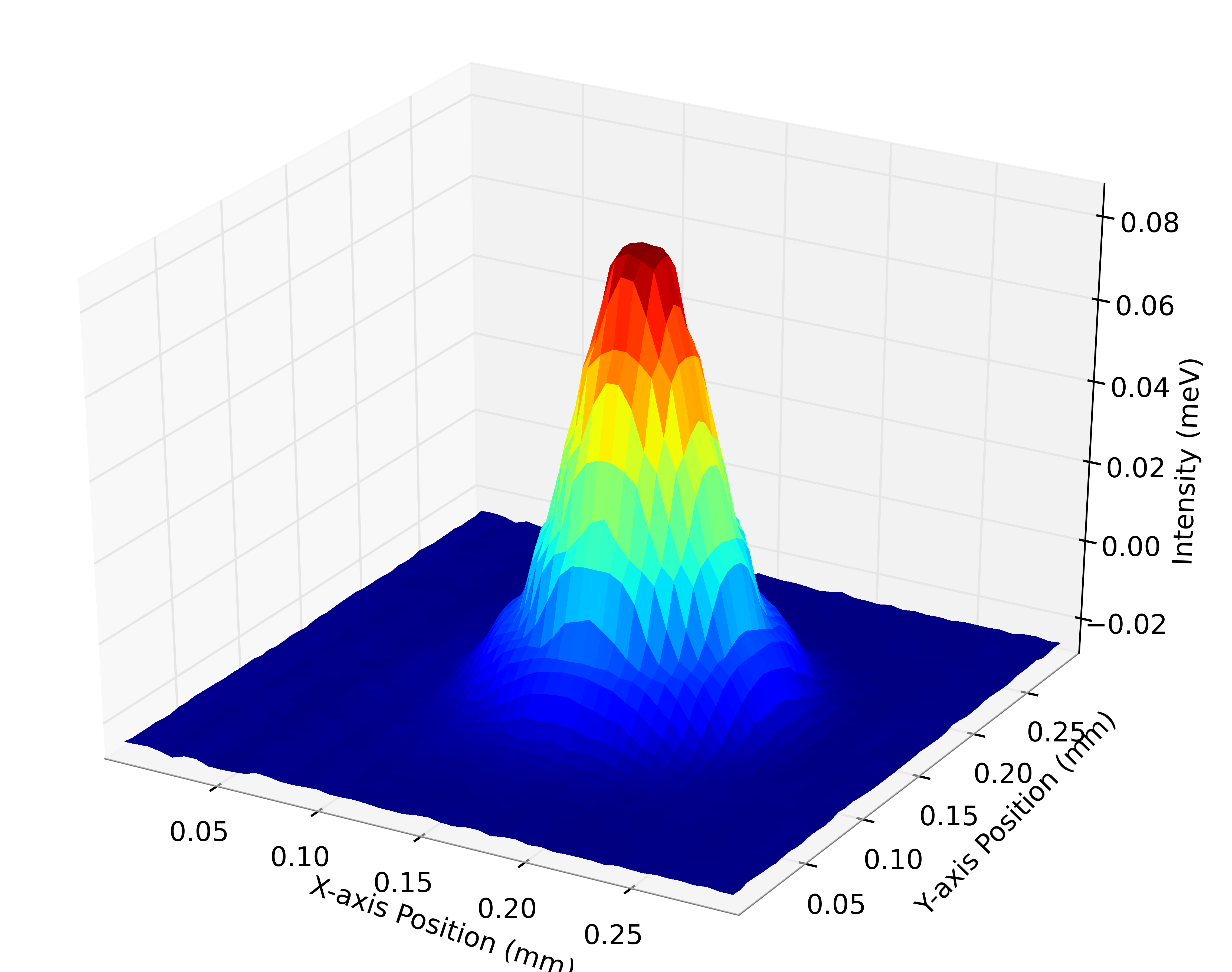}
\includegraphics[width=0.45\linewidth,clip=true]{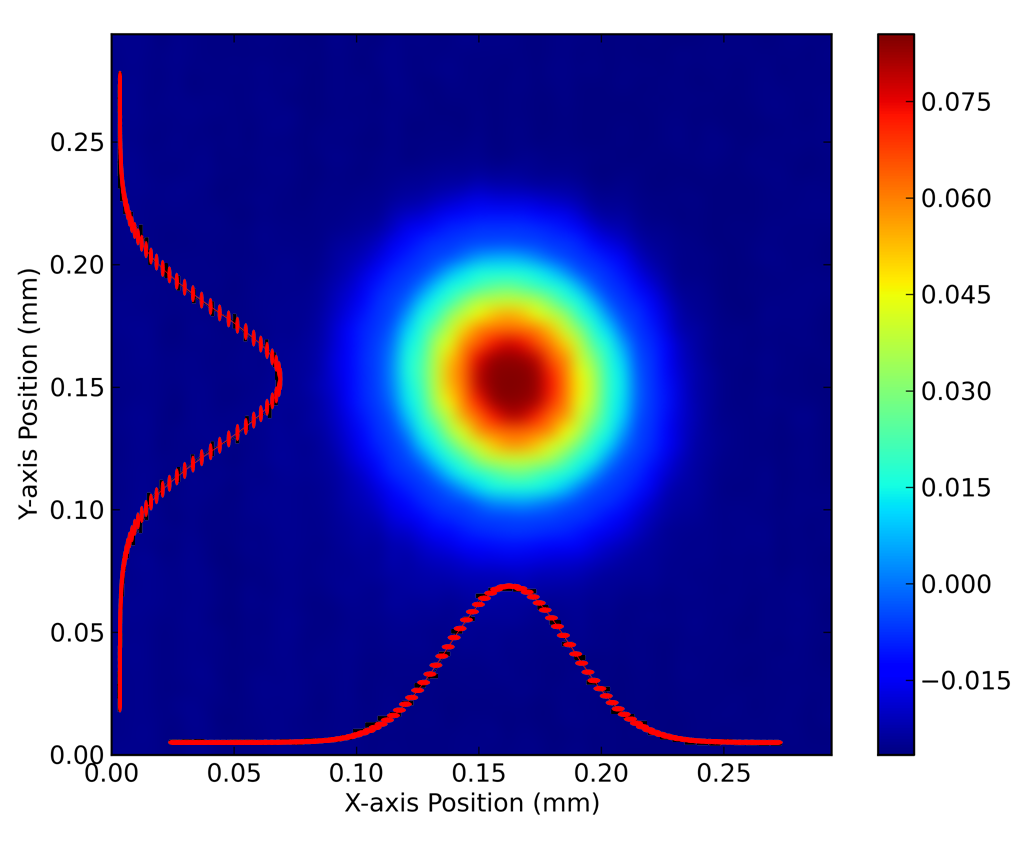}
\caption{Reconstructed intensity distribution profile at $i=0.30 mA$ plotted in 3D (left) and from top (right).  The red lines represent a gaussian fit of the beam shape along two orthogonal radial cuts (along x and y).
}
\label{beam, beamtopquan}
\end{figure}

Two kinds of wavelength information can be obtained from the spectral channel:  peak wavelength and full spectrum.  The peak wavelength is automatically extracted by the optical spectrum analyzer by comparing the intensity value of the maximum in the spectrum to a reference intensity (set at -69 dB).  For all points in the sampled matrix where the peak intensity passes the reference, the optical wavelength and the intensity value are input into the corresponding matrices -- for all other points, no information is output, thus avoiding the collection of noisy data.  In Fig.~\ref{wave0dian19mA,wave0dian2mA,wave0dian3mA} the difference between the measured peak wavelength and a set reference is plotted in a false-color scale for below threshold ($i = 0.19 mA$), threshold ($i = 0.20 mA$) and above threshold ($i = 0.30 mA$).

\begin{figure}[H]
\centering
\includegraphics[width=0.325\linewidth,clip=true]{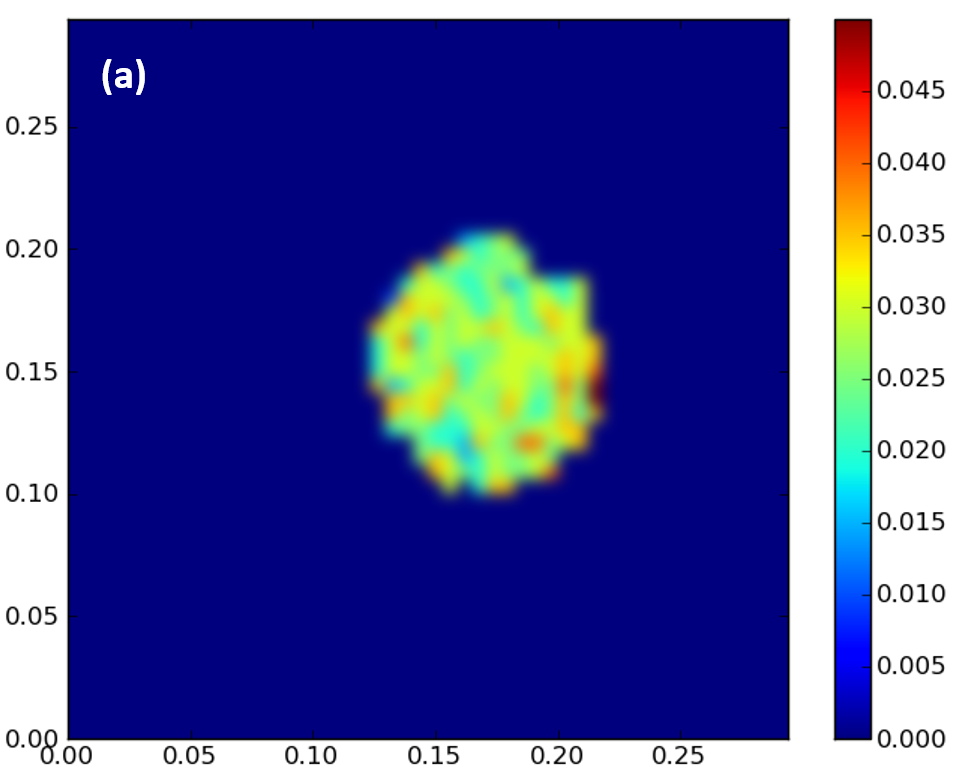}
\includegraphics[width=0.325\linewidth,clip=true]{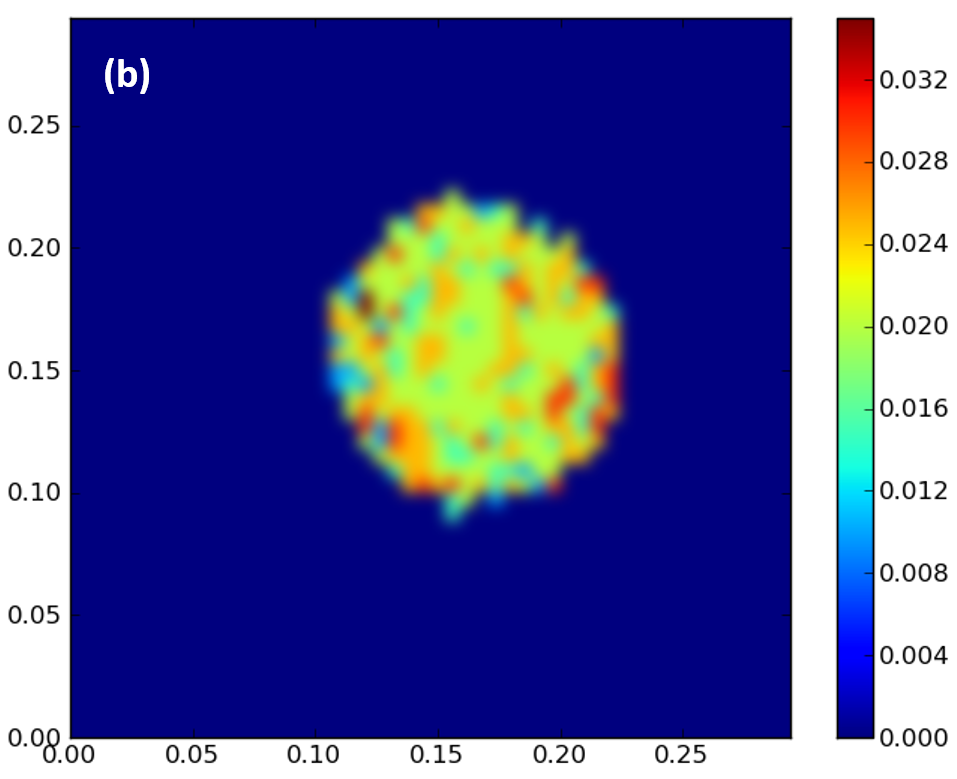}
\includegraphics[width=0.325\linewidth,clip=true]{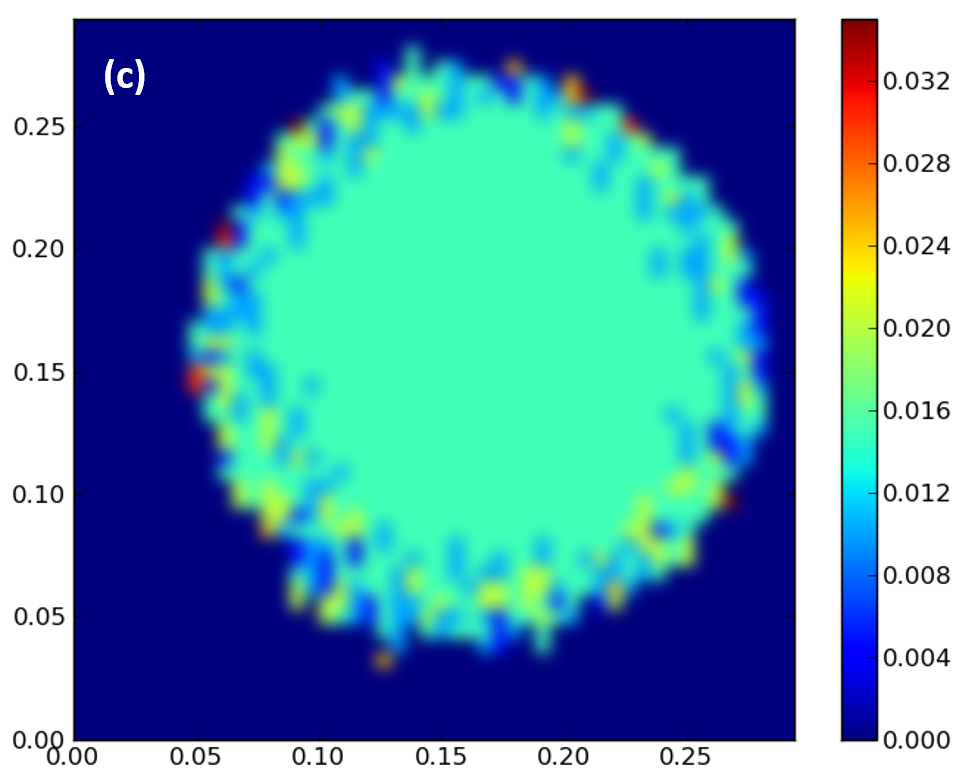}
\caption{Peak wavelength emission distribution across the beam at $i = 0.19$ (a), $0.20$ (b) and $0.30 mA$ (c).  The color scales are adjusted for each figure separately, while a reference frequency is subtracted from each figure. 
}
\label{wave0dian19mA,wave0dian2mA,wave0dian3mA}
\end{figure}

As expected, when the laser is operating below threshold (Fig.~\ref{wave0dian19mA,wave0dian2mA,wave0dian3mA}(a)), or at threshold (Fig.~ \ref{wave0dian19mA,wave0dian2mA,wave0dian3mA}(b)), the wavelength distribution across the beam is inhomogeneous but the interval of spanned wavelengths decreases with increasing current.  When the pumping current is above threshold, e.g. $i = 3.00 mA$, the central part of the beam exhibits a homogeneous emission wavelength. At the same time, the wavelength at the beam's edges still fluctuates by as much as $0.3$\AA.  Comparison to the 3D beam profile representation shows a deviation from the gaussian profile in a ring near the beam's base corresponding to the region where the wavelength fluctuations are found; this is also the area of the device where current crowding occurs, due to the ohmic contact, thus directly imaging the consequence of perturbations on the optical beam.  At the same time, the very low field intensity values in this region of the beam are liable to feel a nonnegligible contribution coming from the spontaneous emission.  The observed wavelength fluctuations are indeed compatible with the filtering action of the cavity (with estimated $\Delta \lambda_c = 1 \ldots 10$ \AA) acting onto the spontaneous line.  This illustrates the power of the simultaneous imaging technique.

\begin{figure}[H]
\centering
\includegraphics[width=0.65\linewidth,clip=true]{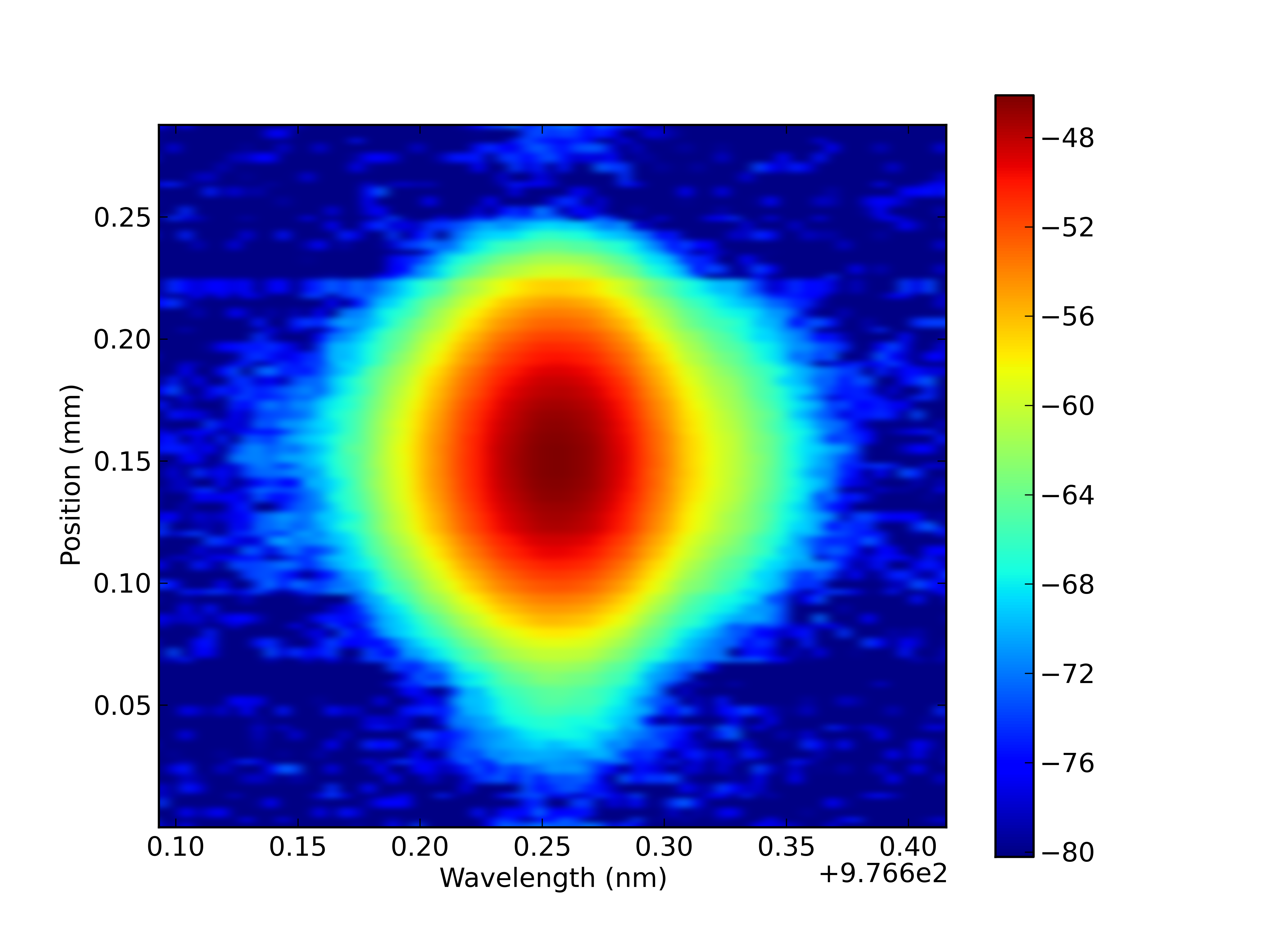}
\caption{Full optical spectrum (horizontal axis) measured along a horizontal beam cut (vertical axis) passing through the beam center.
}
\label{spectrum2}
\end{figure}

Fig.~\ref{spectrum2} shows the entire frequency spectrum (horizontal axis) measured along a horizontal cut passing through the beam center (displayed vertically in the figure) at $i = 0.30 mA$.

\begin{figure}[H]
\centering
\includegraphics[width=0.65\linewidth,clip=true]{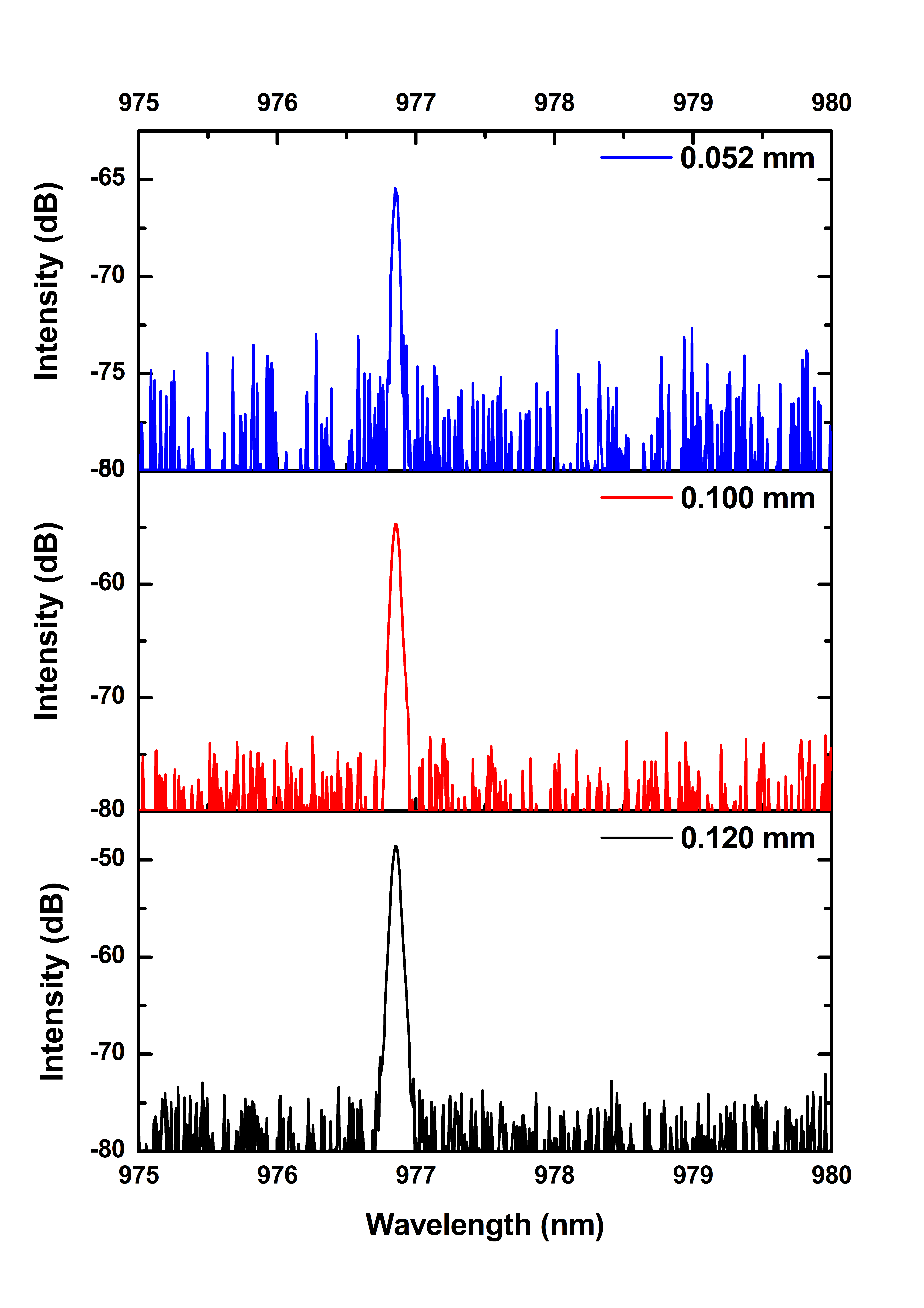}
\caption{Full optical spectrum at selected  positions in the beam (cf. label).  $i = 0.30 mA$.
}
\label{spectrumposition}
\end{figure}

This is the second kind of spectral information which can be gathered from our device, where the full spectrum can be acquired for the whole beam and can be easily displayed along one of the two main directions.  Details of the spectrum at selected positions (cf. caption of Fig.~\ref{spectrumposition}) are plotted in the figure.  Notice that the spectral noise is well below the reference value (-69 dB) chosen for the spectral peak information.

\begin{figure}[H]
\centering
\includegraphics[width=0.65\linewidth,clip=true]{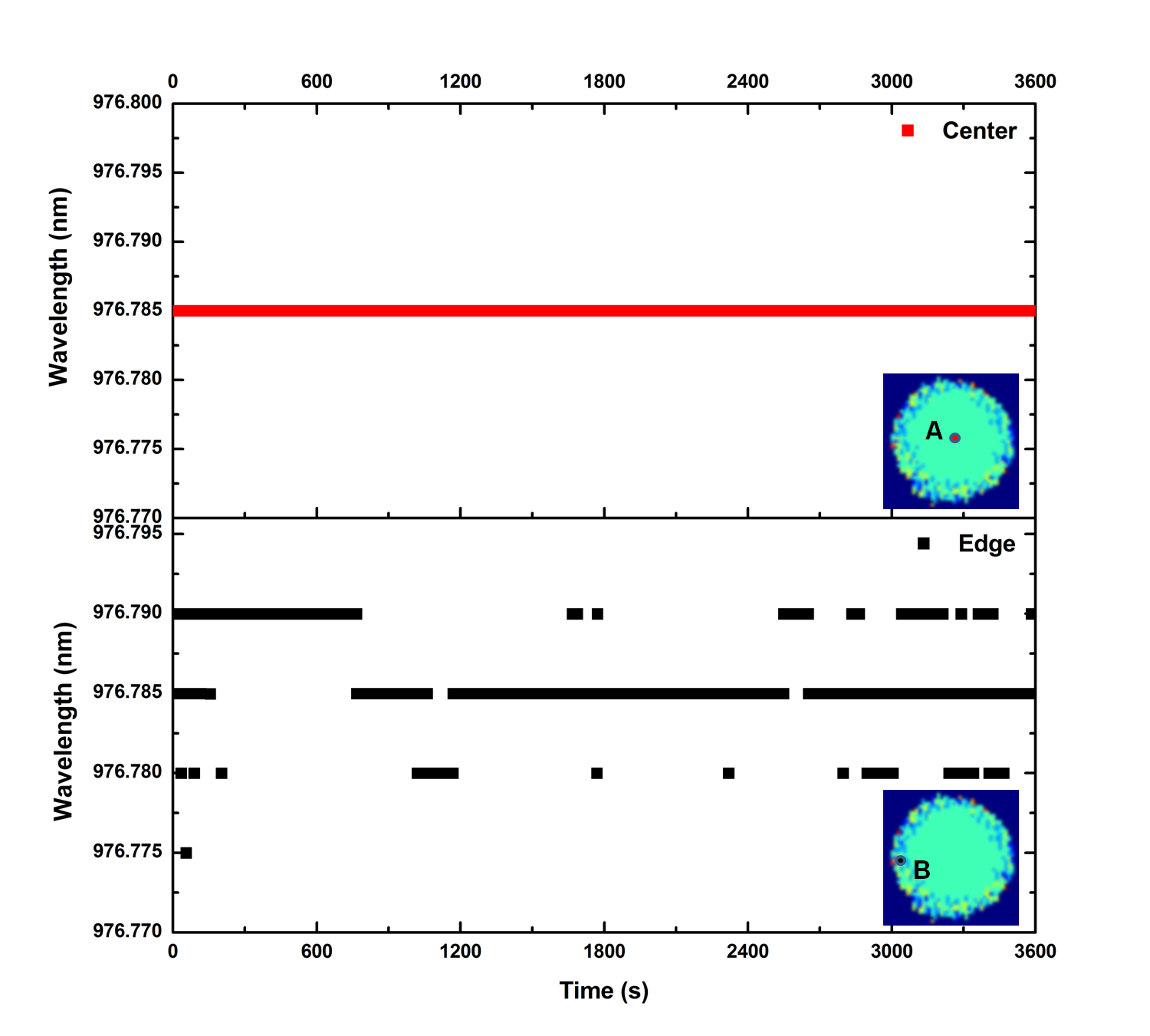}
\caption{Temporal stability of the emission wavelength at two different positions, displayed in each inset.  $i = 0.30 mA$.
}
\label{wavetime}
\end{figure}

The intrinsically slow nature of the scan provides average information on the full scale figure, thus stability issues are important.  Placing the fiber at selected positions (on-axis, Fig.~\ref{wavetime} top; near the beam's edge, Fig.~\ref{wavetime} bottom) allows for stability tests.  While on-axis the wavelength remains perfectly stable over time scales covering the acquisition time --within the optical spectrum analyzer's reproducibility ($0.002 nm$) --, at the beam's edge -- i.e., where fluctuations appear in Fig.~\ref{wave0dian19mA,wave0dian2mA,wave0dian3mA} --, one observes slow temporal wavelength changes.  Thus, we can conclude that the information pertaining to beam's center remains constant over the scan time, while the fluctuations observed around the edges (Fig.~\ref{wave0dian19mA,wave0dian2mA,wave0dian3mA}) must be interpreted as covering the range of possible emission wavelengths occurring in the ring, rather than spatially-dependent, but time-independent, wavelengths associated with each individual measured point.

\section{Conclusions}

Summarizing, we have shown a simultaneous characterization of the beam emitted by a semiconductor microcavity with high spatial and optical resolution.  The single-mode gaussian shaped intensity distribution is accompanied by a homogeneous emission frequency, as expected, but interesting wavelength fluctuations appear in the beam's wings, influenced both by the spontaneous emission and by manufacturing details.

Coupling the intensity channel to a linear detector, as presently reported, allows for a digitization depth which easily surpasses the standard CCD performance, since the dynamic range is solely determined by the digitizer which follows (and which can easily reach 16 bits with low bandpass -- $< 1 Hz$).  Additional dynamic range can be gained with the use of a logarithmic detector, which improves the resolution in the low-intensity regions of the beam and offers additional potential for investigating their details.  This performance is easily matched by the optical spectrum analysis, intrinsically logarithmic, thus allowing for detailed coupled studies of the low-intensity regimes.
Comparing with commercial instruments, our measurement system is flexible, possesses high resolution and offers multichannel information. 

In this paper we have shown an example of laser characterization in the near field, but the far field and any intermediate plane can be easily analyzed by replacing the optical system preceding the fiber's pick-up plane and by suitably adjusting the scanning range.  While we have shown a double-channel measurement (intensity and wavelength), the apparatus can be readily extended to measuring multiple channels including, for instance, local polarization measurements (with adequate fibers), interferometric measurements (e.g., field coherence), e.m. field amplitude and phase measurements~\cite{Kelleher2010} and the observation of fast temporal dynamics. The high resolution possible, both in the near field (shown here) as well as in the far field allows for a detailed characterization of microcavities, of laser beam parameters, of LEDs, of large-diameter optical fibers or light-pipes, and of local defects in light emitting devices, with possible applications to material science, biology, chemistry, etc.

\section{Acknowledgements}
We are grateful to S. Barland, B. Garbin, L. Gil, M. Giudici, F. Gustave, M. Marconi for discussions and loan of instrumentation. Technical support from J.-C. Bery (mechanics) and from J.-C. Bernard and A. Dusaucy (electronics) is gratefully acknowledged.
T.W. acknowledges a Ph.D. Thesis contract from the Conseil R\'egional PACA and support from BBright.

\section{Appendix A}\label{appendixA}
\begin{figure}[H]
\centering
\includegraphics[width=0.6\linewidth,clip=true]{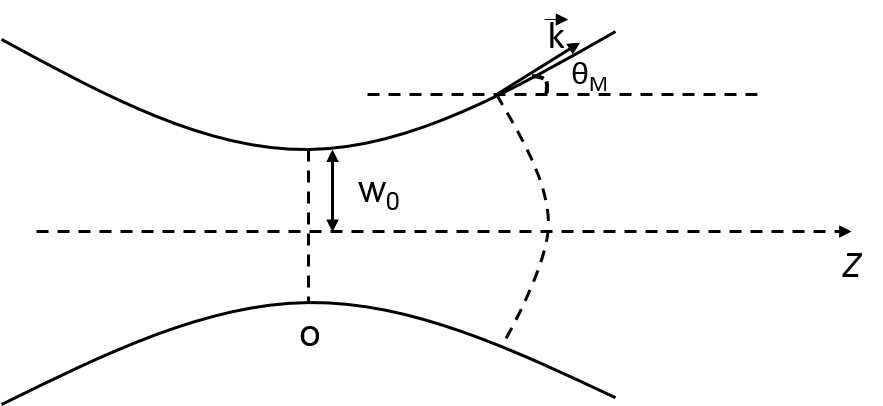}
\caption{
Schematics of a gaussian beam with beam waist $w_0$ at the origin of the reference system ($z = 0)$.  The solid lines denote the envelope of the $(1/e)$ point for the field (i.e., the $(1/e^2)$ for the intensity).  At an arbitrary point $z$ the wavefront is curved and its wavevector $\vec{k}$, forming an angle $\theta_M$ with the optical axis, is explicitely marked in the figure.  Cf. text for details.
}
\label{gaussian}
\end{figure}

Care must be taken in assessing the angle of incidence relative to the fiber across the beam to be sampled.  Indeed, in a generic position along the optical axis, unless the wavefront is planar, the wavevector is going to form an angle with the axis, thus also with the fiber axis which has been previously aligned with the optical axis of the laser.  For illustration we consider an axisymmetric gaussian beam with waist $w_0$ at $z = 0$ (Fig.~\ref{gaussian}).  

For any coordinate $z$ (taken positive to simplify the discussion, since by symmetry the problem is identical on either side of $z = 0$) to the radial position on the locus 
\begin{eqnarray}
w(z) = w_0 \sqrt{1+\left( \frac{z}{z_R} \right)^2}
\end{eqnarray}
corresponds to intensity~\cite{Siegman1986} $I(r) = \frac{I_a}{e^2}$  ($I_a$ being the value of the intensity on axis:  $r=0$).  Since this point lies in the wings of the gaussian (radial) intensity distribution, we will consider it as a reference for our calculation -- any other point, even further out in the wings, can be chosen and the calculation can be repeated along the same lines.

The angle which the wavevector, orthogonal to the wavefront, forms with the optical axis is determined by
\begin{eqnarray}
\tan \theta_M & = & \frac{w(z)}{R(z)} \, , \\
R(z)  & = & z + \frac{z_R^2}{z}
\end{eqnarray}
where $z_R = \frac{\pi w_0^2}{\lambda}$ is the Rayleigh length and $\lambda$ the wavelength of the radiation~\cite{Siegman1986}.  The explicit expression for the angle therefore reads:
\begin{eqnarray}
\tan \theta_M & = & \frac{\lambda}{\pi w_0} \frac{z}{z_R} \frac{1}{\sqrt{1 + \left( \frac{z}{z_R} \right)^2}} \, , \\
& = & \theta_{ff} \frac{\zeta}{\sqrt{1 + \zeta^2}} 
\end{eqnarray}
where we have defined, for simplicity, the normalized $\zeta = \frac{z}{z_R}$ coordinate and $\theta_{ff} \equiv \frac{\lambda}{\pi w_0}$ represents the so-called {\it far field angle} \cite{Siegman1986}, i.e., the angle that the wavevector forms with the optical axis when $\zeta \rightarrow \infty$.  It is immediate to see that 
\begin{eqnarray}
\frac{\zeta}{\sqrt{1 + \zeta^2}} & \rightarrow & 1
\end{eqnarray}
as a monotonic function when $\zeta \rightarrow \infty$, thus, the angle of the wavevector is always $0 \le \theta_M \le \theta_{ff}$. 

In order for the light to be correctly coupled into the fiber, its angle in the beam's wings $\theta_M$ needs to match the angle of acceptance of the fiber $\theta_a$, defined by the Numerical Aperture~\cite{Saleh2007} (NA):
\begin{eqnarray}
\label{defNA}
NA & = & \sin \theta_a \\
& = & \sqrt{n_1^2 - n_2^2}
\end{eqnarray}
with $n_1$ core and $n_2$ cladding refractive index, respectively.

Thus, in order to couple correctly the light into the fiber at all points included in the range, we must achieve the condition
\begin{eqnarray}
\tan \theta_M & \le & \sin \theta_a \, ,\\
\nonumber
& {\rm i.e.,} & \\
\label{cond}
\theta_{ff} \frac{\zeta}{\sqrt{1+ \zeta^2}} & \le & NA
\end{eqnarray}

The above expression holds for any position of observation along the propagation axis $z$.

The most useful planes of observation in an optical system are the near and far field.  The latter is obtained, at least for a semiconductor laser, using a collimator, which transforms the diverging gaussian beam into a (nearly) plane front (cf. text for a discussion).  In the near field, i.e. on the plane of the beam waist ($z = 0$), the wavefront is also planar, thus the constraint of keeping a constant coupling efficiency into the fiber across the whole beam is automatically satisfied.  However, axial positioning errors may entail some deviation from optimum coupling in the beam's wings.  It is easy to use eq.~(\ref{cond}) to estimate the magnitude of the error.

Since we are considering a small error relative to the position of the waist (i.e., $\zeta \ll 1$), we can neglect the quadratic term in the square root and obtain an upper value of the tolerable displacement from the actual $\zeta = 0$ plane:
\begin{eqnarray}
\zeta & \le & \frac{1}{\theta_{ff}} NA \, , \\
\label{Deltaz}
\Delta z & = & \frac{z_R}{\theta_{ff}} NA \, , \\
& \le & \frac{\pi^2 w_0^3}{\lambda^2} NA
\end{eqnarray}

Using a mode field diameter for a standard single-mode fiber $(MFD) = 6.6 \mu m$ and using the definition of the $V$-number, we can estimate the Numerical Aperture of our single-mode fiber:
\begin{eqnarray}
NA & = & \frac{\lambda V}{\pi (MFD)}\\
& \approx & 0.12 \, .
\end{eqnarray}
This allows us to estimate the constraints in the main text.

\section{Appendix B}\label{appendixB}

Because of their practical importance, scanning instruments have been strongly developed in the past few decades and have reached high-level performances.  In order to place our instrument in the context of the existing devices, we compare its characteristics to those of an average instrument available on the market.  We have to remark, however, that the prime feature of our device, which does not exist for the moment in any commercial beam profiler, is the simultaneous measurement of the intensity distribution and its spectral composition (or even the possibility for multiple simultaneous measurements).  In this respect, our instrument is unique.

Commercial beam profilers can be divided into two classes:  camera-based (normally CCD) or knife-edge.  The performances of the two categories are quite different as the former offer a true 2D image, while the latter are based on the integral of the power collected as a knife-edge passes in front of the beam (and require symmetric beams for meaningful measurements).

The spatial resolution of a CCD-based device is normally in excess of one Megapixel where the typical pixel side measures in excess of $10 \mu m$ (special devices with pixel size $5 \mu m$ exist) -- the pixels are normally rectangular.  Our device offers a spatial resolution determined by the optical fiber.  At $\lambda \approx 1 \mu m$ a single mode fiber has a diameter $6 \sim 7 \mu m$ and offers therefore comparable resolution.  However, while the matrix is adaptable and can be chosen to match the beam, acquisition times limit the total number of ``pixels" in our instrument to less than $10^4$ points.  It must be noticed, however, that CCDs have a fixed size and resolution and that often a good part of the chip remains unused, while in our instrument the scanning range can be straightforwardly adapted to each situation, thus optimizing its performance.  This way, the apparently very large reduction in spatial resolution from which our instrument seems to suffer does not correspond to a true limitation in performance.  

CCD-based devices possess a dynamic range which is rather limited ($<$ 8 bits for most cameras, $<$ 12 bits for high performance, thus at best 4000 grey levels -- not counting offset, possible nonlinear regions, and saturation near the upper limit of sensitivity).  Since our instrument makes use of a detector and, due to its long overall measurement time for each point, can use high precision, variable scale converters, its dynamic range can extend over several orders of magnitude (at least 4 or 5).  Use of a logaritmic detector can further improve the dynamic range, rendering its performance unsurpassable by any reasonable CCD-based device.  Knife-edge beam profilers, which also make use of an integrating detector, can compete on this point but, due to their faster response, cannot reach the capabilities of our instrument.  Linearity is an additional concern, since CCD-based devices can easily suffer from a somewhat nonlinear response (due, in part , to pixel-to-pixel cross-talk).  Since measurements are taken serially, our instrument is devoid of cross-talk (as long as the sampling is correctly chosen to avoid overlap between pixels) and offers a linearity as good as that of the detector employed (tipycally far better than the one provided by any CCD camera) and of the Digital-to-Analog-Converter (DAC).

The CCD-based devices have one intrinsic shortcoming which originates from the difficulties arising from a lack of spatial uniformity in the response.  Good performance can be achieved only with very high-end, thus expensive, chips, compensated, in part, by calibrations which need to be run (sometimes automatically, in the background).  The cross-talk, mentioned above, represents an additional factor of distortion from linearity and, unfortunately, leads to intensity-dependent cross-talk.  The typical cross-shaped background enhancement in correspondence of an intense beam (or part thereof) represents and intrinsic shortcoming which can hardly be compensated.  Our instrument, working on a serial acquisition on a single sensitive element, is entirely free from any of these perturbations and guarantees a much better linearity across the whole pattern.

As far as the response time is concerned, our device cannot compare to either CCD- or knife-edge-based devices. Both have (equivalent) frame rates up to 10Hz, while our instrument requires minutes (or hours) for a full image -- the high linearity, high dynamic range, good spatial resolution and cross-talk-free performance offset this shortcoming.  As a matter of fact, the instrument we are presenting can only be used on temporally very stable beams, or, alternately, can only provide statistical information about the beam itself.  As such, unless the conditions are very particular, our multichannel device will hardly be suitable for acquiring profiles on pulsed lasers.

The aim of most beam profilers is to be suited to a range of wavelengths as large as possible.  Any single measurement head (e.g., a CCD) may operate over the visible and near-infrared (up to 1.1 $\mu m$) and instruments with different {\it heads} -- each adapted to a wavelength interval (up to 16 $\mu m$ for some heads) -- exist.  The same applies to our instrument which requires different optical fibers and (possibly) different detectors, but the costs associated with the changes are definitely smaller than those of replacing a CCD.

CCD-based devices advertise sensitivities as low as 50$\mu W cm^{-2}$.  With the single-mode fiber and the UDT-455 detector (cf. Section~\ref{expt}), with sensitivity approaching $10^7 V/W$ we can nearly match this performance ($300 \mu W cm^{-2}$ in this configuration, assuming a $mV$ measurement resolution).  Since a slow DAC converter can easily offer better resolutions (at least one order of magnitude -- not counting {\it nanovoltmenters} offering $1 \mu V$ and improving the performance by three orders of magnitude), the sensitivity can be improved well beyond the current performances of commercial CCD-based beam profilers.


\begin{thebibliography}{40}

\bibitem{Demtroder2009}W. Demtr\"oder, Laser Spectroscopy, 3$^{\rm rd}$ ed. (Springer, Berlin, 2009).
\bibitem{Duarte2009} F. J. Duarte, Tunable Laser Applications, 2$^{\rm nd}$ ed. (CRC, Boca Raton, FL, 2009).
\bibitem{Sparkes2008}M. Sparkes, M. Gross, S. Celotto, T. Zhang, W. O'Neil, J. Laser Appl., {\bf 20}, 59-67 (2008).
\bibitem{Gavrilovic1995}P. Gavrilovic, M. Wober, K. Meehan, and M. S. O'Neill, IEEE J. Quantum Electron., \textbf{31}, 623-626 (1995).
\bibitem{Sheikh2011} M. Sheikh, P. Marraccini, and N. A. Riza, %Laser Beam Characterization using Agile Digital-Analog Photonics.
Proc. SPIE, {\bf 7675}, 767508 (1-6) (2011).
\bibitem{Duocastella2012}M. Duocastella and C. B. Arnold, %Bessel and annular beams for materials processing.
Laser Photonics Rev. {\bf 6}, 607-621 (2012).
\bibitem{Gan2013}Z. S. Gan, Y. Y. Cao,	R. A. Evans	and M. Gu,
%Three-dimensional deep sub-diffraction optical beam lithography with 9nm feature size.
Nature Commun., {\bf 4}, 2061, (2013).
\bibitem{Schulze2010} M. Schulze, %Medical Applications of Lasers: Diversity is Key to Success.
Laser in der Medizin, {\bf 6}, 32-35 (2010).
\bibitem{Kataoka2001} K. Kataoka, Y. Shibayamai and K. Doi, %Multiple beam scanning optics for laser printer application of optical Fiber array method.
Optical Review {\bf 8}, 218-226, (2001).
\bibitem{Holmes2006} A. S. Holmes, J. E. A. Pedder and K. L. Boehlen, %Advanced Laser Micromachining Processes for MEMS and Optical Applications.
Proc. SPIE {\bf 6261}, 62611E (1-9) (2006)
\bibitem{Parvin2004} P. Parvin, B. Sajad, K. Silakhori, M. Hooshvar, and Z. Zamanipour,%Molecular Laser Isotope Separation versus Atomic Vapor Laser Isotope Separation, 
Prog. Nuclear En. {\bf 44}, 331-345 (2004), and references therein.
\bibitem{Minoshima2001} K. Minoshima, A.M. Kowalevicz, I. Hartl, E.P. Ippen, and J.G. Fujimoto, %Photonic device fabrication in glass by use of nonlinear materials processing with a femtosecond laser oscillator,
Opt. Lett. {\bf 26}, 1516 (2001).
\bibitem{Ready2001} J.F. Ready, D.F. Farson, T. Feeley (Eds.), LIA Handbook of Laser Materials Processing, (Springer, Berlin, 2001).
\bibitem{Steen2010} W. Steen, K.G. Watkins, J. Mazumder, Laser Material Processing, (Springer, Berlin, 2010)
\bibitem{Ashkin2006} A. Ashkin, Optical Trapping and Manipulation of Neutral Particles Using Lasers, (World Scientific, Singapore, 2006).
\bibitem{Dickey2005} F. M. Dickey, S. C. Holswade, and D. L. Shealy, eds., Laser Beam Shaping Applications (CRC Press, 2005).
\bibitem{Siegman1998} A.E. Siegman, How to (Maybe) Measure Laser Beam Quality, in ``DPSS (Diode Pumped Solid State) Lasers: Applications and Issues", M. Dowley, ed., Vol. 17 of OSA Trends in Optics and Photonics p. 184 (paper MQ1).  (Optical Society of America, 1998).
\bibitem{Roundy} C.B. Roundy, Current technology of laser beam profile measurements, in ``Laser Beam Shaping: Theory and Techniques", F.M. Dickey and S.C. Holswade, eds., 349-420 (CRC Press, Boca Raton, FL, 2002).
%\bibitem{Takemura2012} N. Takemura, J. Omachi and M. Kuwata-Gonokami, Fast periodic modulations in the photon correlation of single-mode vertical-cavity surface-emitting lasers. Phys. Rev. A {\bf 85}, 1-5, (2012).
\bibitem{Weber1992} H. Weber, %Some historical and technical aspects of beam quality,
Optical Quantum Electron., {\bf 24}, 861-864 (1992).
\bibitem{Siegman1993} A.E. Siegman, %Defining, measuring, and optimizing laser beam quality, in
%``Laser Resonators and Coherent Optics: Modeling, Technology, and Applications", A. Bhowmik ed.,
Proc. SPIE {\bf 1868}, 2 (1993).
\bibitem{Alsultanny2006} Y. A. Alsultanny, Journal of Computer Science \textbf{2}, 109-113 (2006)
\bibitem{Gebert2014} F. Gebert, M. H. Frosz, T. Weiss, Y. Wan, A. Ermolov, N. Y. Joly, P. O. Schmidt, and P. St. J. Russell, Optics Express \textbf{22}, 15388-15396 (2014).
\bibitem{Look2010}J.-R. van Look, S. Einfeldt, O. Krüger, V. Hoffmann, A. Knauer, M. Weyers, P. Vogt, and M. Kneissl, IEEE Phot. Technol. Lett. \textbf{22}, 416 (2010)

\bibitem{Liu2011} H. Y. Liu, T. Wang, Q. Jiang, R. Hogg, F. Tutu, F. Pozzi and A. Seeds, Nature Photonics \textbf{5}, 416-419 (2011).
%\bibitem{Heard1968} H.G. Heard, Laser Parameter Measurements Handbook (Wiley, 1968).


\bibitem{ULMPhot} ULM Photonics Single Mode (980$\pm$ 3) nm TO46 \& TEC manufacturer's specifications.

\bibitem{Kelleher2010} B. Kelleher, D. Goulding, B. Baselga Pascual, S.P. Hegarty, and G. Huyet, Eur. Phys. J. D {\bf 58}, 175-179 (2010).

\bibitem{Siegman1986} A. E. Siegman, Lasers, (University Science Books, Mill Valley, CA 1986).

\bibitem{Saleh2007} B.E.A. Saleh and M.C. Teich, Fundamentals of Photonics, 2$^{\rm nd}$ ed., (Wiley, New York, NY, 2007).


\end{thebibliography}
\end{document}